\documentclass[prd,showpacs,twocolumn,amssymb]{revtex4}
\begin{document}

\title{Photons and Fermions in Spacetime with a Compactified Spatial Dimension}
\author{Efrain J. Ferrer}
\author{Vivian de la Incera}
\affiliation{Physics Department, State University of New York at
Fredonia, Fredonia, NY 14063, USA}

\begin{abstract}
The effects of a nonsimply connected spacetime with the topology
of $S^{1}\times R^{3}$ in the vacua of QED and gauged-NJL theories
are investigated. It is shown that the polarization effects of
twisted and untwisted fermions in QED are equivalent, once the
corresponding stable vacuum solution of each fermion class is
taken into account. The photon propagation in QED is found to be
anisotropic and characterized by several massive photon modes and
a superluminal transverse mode. At small compactification radius
the masses of the massive modes increase as the inverse of the
radius, while the massless photon mode has a superluminal velocity
that increases logarithmically with that distance. At low energies
the photon masses lead to an effective confinement of the gauge
fields into a $(2+1)-$dimensional manifold transverse to the
compactified direction. In the gauged-NJL model, it is shown that
for both twisted and untwisted fermions, the smaller the
compactification radius, the larger the critical four-fermion
coupling needed to generate a fermion-antifermion chiral symmetry
breaking condensate. 
\end{abstract}
\maketitle

\section{Introduction}

A significant effort in the recent history of physics bears on the
quest to unify all the known fundamental interactions. This old
Einstein's dream found a partial realization in the Standard
Model, where the electromagnetic, weak and strong nuclear forces
were unified in a single one. The fourth fundamental interaction,
gravity, has been an obstacle in that path. Quantum physics in the
realm of the Riemann space, which is the natural habitat for
gravity, gives rise to an anomalous behavior for the quantum
processes where the gravitational quantum particle, the graviton,
participates. An incomplete unified scenario, where the four
fundamental forces cannot be reduced to a single one, limits the
possibility to understand the physics of our universe at the
earliest times.

An innovative approach in the search for a compatible way to unify
gravity with the other forces was attempted in the early works of
Kaluza and Klein \cite{K-K}. There, the possibility of a physical
universe with extra dimensions was considered. In their approach,
the extra dimensions were compactified in small distances that are
hidden to low energy physics, but manifested as gauge fields in
the remnant flat four-dimensional spacetime. More recently, the
idea of extra compactified dimensions has been extensively used in
supergravity, superstrings and brane theories \cite {Candelas}. In
a different development \cite{Dimopoulos}, extra dimensions have
been also applied at lower energy scales in an attempt to
understand hierarchical scales existing between the weak and
Planck energies. The main new ingredient of this approach is that
only gravitons can propagate in the bulk corresponding to the
extra dimensions, while the other gauge fields of the Standard
Model are constrained to the four-dimensional wall.

Motivated by the importance of the Kaluza-Klein scenario, the
study of QFT in nontrivial spacetime has been the focus of
attention of many investigators in recent years. It is well known
that the global properties of the spacetime, even if it is locally
flat, can give rise to new physics. A seminal discovery in this
direction is the so called Casimir effect \cite{Casimir}. In this
phenomenon, an attractive force appears between neutral parallel
perfectly conducting plates. The materialized attractive force is
mediated by the zero-point fluctuations of the electromagnetic
field in vacuum. Hence, the Casimir force is interpreted as a
macroscopic manifestation of the vacuum structure of the quantized
fields in the presence of domains restricted by boundaries or
nontrivial topologies \cite {Milton}.

As known, QFT in spacetime with non-trivial topology has
nonequivalent types of fields with the same spin \cite{Isham}. The
allowed number of distinct field configurations is determined by
the topological structure of the spacetime; generally being more
than one in non-simply connected spaces. In particular, for a
fermion system in a space-time which is locally flat but with
topology represented by the domain $S^{1}\times R^{3}$ (i.e. a
Minkowskian space with one of the spatial dimensions compactified
in a circle $S^{1}$ of finite length $a$), the non-trivial
topology is transferred into boundary conditions for fermions that
are either periodic (untwisted) or antiperiodic (twisted)

\begin{equation}
\psi \left( t,x,y,z-a/2\right) =\pm \psi \left( t,x,y,z+a/2\right) ,
\label{1}
\end{equation}
while for vector fields, only untwisted configurations are
allowed. In (\ref{1})
the compactified dimension with length $a$ has been taken along the $%
\mathcal{OZ}$-direction.

Quantum electrodynamics (QED) with photons coupled to untwisted
fermions or to a combination of twisted and untwisted fermions is
an unstable theory \cite{Ford}. The instability arises due to
polarization effects of untwisted electrons which produce
tachyonic electromagnetic modes \cite {Ford}. For self-interacting
scalar fields the space periodicity can also produce instabilities
causing a symmetry breaking that makes the massless field to
become massive \cite{Toshimura}. The acquired mass depends on the
periodicity length, so this phenomenon is called topological mass
generation.

To understand in qualitative terms how the fermion boundary
conditions in a non-trivial topology can produce instabilities in
QED, we should have in mind that, thanks to the vacuum
polarization, the photon exists during part of the time as a
virtual $e^{+}e^{-}-pair$. The virtual pair can then transfer the
properties of the quantum vacuum, which as known, depend on the
non-trivial topology and boundary conditions of the space under
consideration, to the photon spectrum.

Our main goal in the present report is to analyze the consequences
of the non-trivial topology for photon propagation in QED and for
fermion condensation in a gauged-NJL theory. We will show that the
non-simply connected character of the spacetime may give rise to
different photon modes of propagation, which are normally absent
in QED in a flat space with trivial topology. Another topic that
we will discuss is how the compactified dimension influences the
chiral symmetry breaking in a gauged-NJL theory.

\section{Non-trivial Vacuum Solutions in Compactified QED}

The vacuum polarization in the non-trivial spatial topology $S^{1}\times
R^{3}$ can be influenced by both virtual untwisted and twisted $%
e^{+}e^{-}-pairs$. The results for twisted fermions can be easily read off
the results at finite temperature, since in the Euclidean space the two
theories are basically the same after the interchange of the four-space
subindexes $3\leftrightarrow 4$. Nevertheless, a different situation occurs
with untwisted fermions that has no analogy in the statistical case.
Henceforth, we concentrate our attention in the untwisted-fermion case.

Let us consider the QED action in a spacetime domain with compactified
dimension of length $a$ in the $\mathcal{OZ}$-direction

\begin{equation}
\mathcal{S}=\int\limits_{-a/2}^{a/2}dx_{3}\int\limits_{-\infty }^{\infty
}dx_{0}d^{2}x_{\bot }\left[ -\frac{1}{4}F_{\mu \nu }^{2}+\overline{\psi }%
(i\partial \llap / -eA\llap / -m)\psi \right] .\qquad  \label{2}
\end{equation}

When this compactified QED action is considered for untwisted
fermions, the effect of vacuum polarization upon photon
propagation gives rise to a tachyonic mass for the third component
of the photon field \cite{Ford}. In a quantum theory the existence
of tachyonic modes are an indication that the considered vacuum is
not the physical one, and that a symmetry breaking mechanism is in
order. Indeed, in compactified QED with untwisted fermions it has
been shown \cite{Romeo} that a constant expectation value of the
electromagnetic potential component along the compactified
direction minimizes the effective potential, thereby stabilizing
the theory. The same stable vacuum solution obtained in QED with
$S^{1}\times R^{3}$ topology in Ref. \cite{Romeo}, is also present
in the case of massless QED with periodic fermions on a circle \cite{Hosotani} (QED with $%
S^{1}\times R^{1}$ topology). Notice that, even though a constant
vacuum configuration has $F_{\mu \nu }=0$, it cannot be gauged to
zero, because the gauge transformation that would be needed does
not respect the periodicity of the function space in the
$S^{1}\times R^{3}$ domain. This is a sort of Aharonov-Bohm effect
that makes $A_{\mu }$ a dynamical variable due to the non-simply
connected topology of the considered spacetime. The lack of gauge
equivalence between a constant component of the gauge potential
($A_{0}$ in this case) and zero is also manifested in QED at
finite temperature and/or density due to the compactification of
the time coordinate \cite{Batakis}.
In the statistical case, however, the minimum of the potential is at $A_{0}=0$%
, since only twisted fermions are allowed. On the other hand, in
the electroweak theory with a finite density of fermions, a
non-trivial constant vacuum $A_{0}$ is induced by the fermion
density and cannot be gauged away \cite{Linde}. There, in contrast
to the system considered in the present paper, an additional
parameter (a leptonic and/or baryonic chemical potential) is
needed to trigger the non-trivial constant minimum for $A_{0}$.

To find the physical vacuum that stabilizes the untwisted fermion
theory, we propose, following Ref. \cite{Romeo}, the following
ansatz \cite{Note1}

\begin{equation}
\overline{A}_{\nu }=\Delta \delta _{\nu 3},  \label{20}
\end{equation}
with $\Delta $ an arbitrary constant that will be determined from
the minimum equation of the effective potential. Due to the
periodicity of the $A_{\mu}$ fields in the $S^{1}\times R^{3}$
space, the gauge transformations $A_{\mu }\rightarrow A_{\mu
}-\frac{1}{e}\partial _{\mu }\alpha $ are restricted to those
satisfying $\alpha (x_{3}+a)=$ $\alpha (x_{3})+2l\pi ,$ $l\in
\emph{Z}$  \cite{Batakis}. Thus, the gauge transformation $\alpha
(x)=(x\cdot n)e\Delta $, which connects the constant field
configuration (\ref{20}) with zero, could not satisfy the required
periodicity condition unless $\Delta $ were given by

\begin{equation}
\Delta =\frac{2l\pi }{ea},\qquad l\in \emph{Z}  \label{200}
\end{equation}

Let us consider then the one-loop effective potential of the theory (\ref{2}%
) around the vacuum configuration (\ref{20})

\begin{equation}
V=-\frac{1}{2}a^{-1}\ln \left( Det\,\overline{G}^{-1}\right).
\label{20a}
\end{equation}
Here $Det\,\overline{G}^{-1}=$ $%
\sum\hspace{-0.4cm}\int%
d^{4}p\det \,\overline{G}^{-1}=\sum\limits_{p_{3}}\int d^{3}p\det \,%
\overline{G}^{-1},$ with $p_{3}=2n\pi /a,$ $(n=0,\pm 1,\pm 2,...)$
being the discrete frequencies associated with periodic fermions.
In (\ref{20a}) $\overline{G}^{-1}=\gamma \cdot \overline{p}+m $ is
the fermion inverse Green's function in the background $\Delta $,
with $\overline{p}_{\mu }=\left( p_{0},\mathbf{p}_{\perp
},p_{3}-e\Delta \right) $.

After the Wick rotation to Euclidean space and summing in $p_{3}$
we obtain

\begin{equation}
V(\Delta )=-4\int\limits_{-\infty }^{\infty
}\frac{d^{3}\widehat{p}}{\left( 2\pi \right) ^{3}}\left[
\frac{\varepsilon _{p}}{2}+a^{-1}{\textit{Re}}\ln \left(
1-e^{-a\left( \varepsilon _{p}-ie\Delta \right) }\right) \right],
\label{20b}
\end{equation}
where $d^{3}\widehat{p}=idp_{4}d^{2}p_{\bot }$ and $\varepsilon
_{p}=\sqrt{\widehat{p}^{2}+m^{2}}$\noindent .

The extremum of the renormalized effective potential satisfies

\begin{eqnarray}\label{25a}
\left. \frac{\partial V(\Delta )}{\partial \Delta }\right|
_{\Delta
=\Delta _{0}}=\qquad\qquad\nonumber\\=-\int\limits_{-\infty }^{\infty }\frac{d^{3}\widehat{p}}{%
\left( 2\pi \right) ^{3}}\frac{2e^{-a\varepsilon _{p}}\sin \left(
ae\Delta _{0} \right) }{1+e^{-2a\varepsilon
_{p}}-2e^{-a\varepsilon _{p}}\cos \left( ae\Delta _{0} \right)
}=0,
\end{eqnarray}

The extrema of Eq. (\ref{25a}) are $\Delta _{0} =\frac{l\pi
}{ea}$, $l\in \emph{Z}$. Nevertheless, the minimum condition
$\partial ^{2}V(\Delta _{\min })/\partial ^{2}\Delta >0$ is only
fulfilled by the subset

\begin{equation}
\Delta _{\min }=\frac{\left( 2l+1\right) \pi }{ea},\qquad l\in
\emph{Z} \label{26a}
\end{equation}

The elements in the set of minima (\ref{26a}) are gauge
equivalent, since they are all connected by allowed gauge transformations ($%
\alpha (x_{3}+a)=$ $\alpha (x_{3})+2l\pi $). It should be pointed
out, however, that the solutions (\ref{26a}) are not gauge
equivalent to the trivial vacuum $\Delta =0,$ since none of them
satisfies (\ref{200}). That is, the trivial vacuum belongs to a
different gauge class.

Substituting with the minimum solution (\ref{26a}) on Eq.
(\ref{20b}) we obtain

\begin{equation}
V(\Delta _{\min })=-4\int\limits_{-\infty }^{\infty }\frac{d^{3}\widehat{p}}{%
\left( 2\pi \right) ^{3}}\left[ \frac{\varepsilon
_{p}}{2}+a^{-1}{\textit{Re}}\ln \left( 1+e^{-a \varepsilon
_{p}}\right) \right] . \label{26b}
\end{equation}
The expression (\ref{26b}) coincides with the one-loop effective
potential of the theory (\ref{2}) for twisted fermions. As
expected, in the $am\ll 1$ approximation, the effective potential
(\ref{26b}) reduces to

\begin{equation}
V(\Delta _{\min })=-\frac{7\pi ^{2}}{360a^{4}},  \label{26aa}
\end{equation}
which is the result reported for twisted fermions in Ref. \cite{Ford}$.$
Thus, the vacuum energy of both classes of fermions coincides if the
corresponding correct vacuum solution is used.

Notice that the parameter $%
\Delta $ only appears in the distribution functions associated to
the sums in $p_{3}$. Therefore, when $\Delta $ is evaluated at the
minimum solution (\ref{26a}), it turns the statistics of the
untwisted fermions into that of the twisted ones.

\section{Photon Progatation in $S^{1}\times R^{3}$ Spacetime}

\medskip To study the photon propagation in the $S^{1}\times R^{3}$ domain, we should solve the dispersion
relations of the electromagnetic modes, which in the low-frequency
limit has the general form

\begin{equation}
k_{0}^{2}-\mathbf{k}^{2}+\Pi (\mathbf{k}^{2})=0,  \label{2a}
\end{equation}
where $\Pi (\mathbf{k}^{2})$ accounts for vacuum polarization effects.
Different external conditions, as external fields, geometric boundary
conditions, temperature, etc., may modify the vacuum and produce, through $%
\Pi (\mathbf{k}^{2})$, a variation in the spectrum of the photon modes.

The solution of the photon dispersion equations (\ref{2a}) can be
obtained as the poles of the photon Green's function. Due to the
explicit Lorentz symmetry breaking in the $S^{1}\times R^{3}$
topology, we must consider, in addition to the usual tensor
structures $k_{\mu }$ and $g_{\mu \nu }$, a spacelike unit vector
pointing along the compactified direction $n^{\mu }=(0,0,0,1)$.
Then, the general structure of the electromagnetic field Green's
function is

\begin{eqnarray}\label{8}
\Delta _{\mu \nu }(k)=P(g_{\mu \nu }-\frac{k_{\mu }k_{\nu }}{k^{2}})+\qquad\qquad\nonumber\\+Q[\frac{%
k_{\mu }k_{\nu }}{k^{2}}-\frac{k_{\mu }n_{\nu }+n_{\mu }k_{\nu }}{(k\cdot n)}%
+\frac{k^{2}n_{\mu }n_{\nu }}{(k\cdot n)^{2}}]+\frac{\alpha
}{k^{4}}k_{\mu }k_{\nu },
\end{eqnarray}
where $\alpha $ is a gauge fixing parameter corresponding to the
covariant
gauge condition $\frac{1}{\alpha }\partial _{\mu }A_{\mu }=0$, and $P$ and $%
Q $ are defined as

\begin{eqnarray}\label{9}
P=\frac{1}{k^{2}+\Pi _{0}},\qquad\qquad\qquad\qquad\qquad
\nonumber\\Q=-\frac{\Pi _{1}}{(k^{2}+\Pi _{0})\left\{ k^{2}+\Pi
_{0}-\Pi _{1}[k^{2}/(k\cdot n)^{2}+1]\right\} }.\qquad
\end{eqnarray}
The parameters $\Pi _{0}$ and $\Pi _{1}$ are the coefficients of
the polarization operator $\Pi _{\mu \nu }$, which in the
$S^{1}\times R^{3}$ space can be written as

\begin{eqnarray}\label{10}
\Pi _{\mu \nu }(k)=\Pi _{0}(g_{\mu \nu }-\frac{k_{\mu }k_{\nu
}}{k^{2}})+\qquad \nonumber\\+\Pi
_{1}[\frac{k_{\mu }k_{\nu }}{k^{2}}-\frac{k_{\mu }n_{\nu }+n_{\mu }k_{\nu }}{%
(k\cdot n)}+\frac{k^{2}n_{\mu }n_{\nu }}{(k\cdot n)^{2}}].
\end{eqnarray}

From (\ref{8})-(\ref{9}) the photon dispersion relations are

\begin{equation}
k_{0}^{2}-\mathbf{k}^{2}+\Pi _{0}=0,  \label{11}
\end{equation}

\begin{equation}
k_{0}^{2}-\mathbf{k}^{2}+(\Pi
_{0}-\frac{\widehat{k}^{2}}{k_{3}^{2}}\Pi _{1})=0,  \label{12}
\end{equation}
with $\widehat{k}^{2}=k_{0}^{2}-k_{\bot }^{2}$ and $k_{\bot
}^{2}=k_{1}^{2}+k_{2}^{2}$. We point out that in addition to the
transverse mode associated to Eq. (\ref{11}) (normally present in
Minkowski spacetime with trivial topology), a longitudinal mode,
Eq. (\ref{12}), arises here due to the presence of the extra
coefficient $\Pi _{1}$. The situation resembles the finite
temperature case. Nevertheless, as discussed below, the physical
consequences of the spatial compactification are radically
different from those at finite temperature.

The compactified $\mathcal{OZ}$-direction distinguishes itself
from the other spatial directions, so it is convenient to separate
the analysis between photons propagating along $\mathcal{OZ}$
($k_{\perp }=0$), and photons propagating perpendicularly to that
direction ($k_{3}=0$).

The dispersion relations (\ref{11}) and (\ref{12}) for photons propagating
perpendicularly to the compactified direction ($k_{3}=0$) are found, from
Eqs.(\ref{10})-(\ref{12}), to reduce respectively to

\begin{equation}
k_{0}^{2}-k_{\perp }^{2}-\frac{\widehat{k}^{2}}{k_{\bot }^{2}}\Pi
_{00}=0, \label{13}
\end{equation}

\begin{equation}
k_{0}^{2}-k_{\perp }^{2}-\Pi _{33}=0.  \label{14}
\end{equation}

To find the solutions of Eqs. (\ref{13})-(\ref{14}) at the
one-loop level, we need to calculate the one-loop polarization
operator components $\Pi _{00}$ and $\Pi _{33}$ for untwisted
fermions. Considering the free propagator of untwisted fermions at
the minimum solution (\ref{26a})

\begin{equation}
\widetilde{G}(x-x^{\prime })=\frac{1}{(2\pi )^{3}a}%
\sum\hspace{-0.5cm}\int%
d^{4}p\exp [ip\cdot (x-x^{\prime })]G(\widetilde{p}),  \label{14a}
\end{equation}
where

\begin{equation}
G(\widetilde{p})=\frac{\widetilde{p}\llap / -m}{\widetilde{p}%
^{2}-m^{2}+i\epsilon },\qquad \widetilde{p}_{\mu }=\left(
p_{0},p_{\perp },p_{3}-e\Delta _{\min } \right),  \label{4}
\end{equation}
and $%
\sum\hspace{-0.4cm}\int%
d^{4}p=\sum\limits_{p_{3}}\int d^{3}p,$ $p_{3}=2n\pi /a,$ $(n=0,\pm 1,\pm
2,...)$ being the discrete frequencies associated to periodic fermions, the
corresponding one-loop polarization operator is given by

\begin{eqnarray}\label{15}
\Pi _{\mu \nu }(k)=\qquad \qquad \qquad \qquad \qquad \qquad \qquad \qquad \qquad \qquad \nonumber\\=-\frac{4ie^{2}}{(2\pi )^{3}a}%
\sum\hspace{-0.5cm}\int%
d^{4}p\{\frac{\widetilde{p}_{\mu }(\widetilde{p}_{\nu }-k_{\nu })-\frac{1}{2}[\widetilde{p}\cdot (\widetilde{p}%
-k)-m^{2}]g_{\mu \nu }}{(\widetilde{p}^{2}-m^{2})\left[ (%
\widetilde{p}-k)^{2}-m^{2}\}\right]
}\nonumber\\+\mu\leftrightarrow\nu\}\qquad \qquad \qquad\qquad
\qquad.
\end{eqnarray}

In the $a\left| \widehat{k}\right| \ll am\ll 1$ limit, we obtain

\begin{eqnarray}\label{28}
\Pi _{00}(k_{3}=0,k_{0}=0,k_{\bot }\sim 0)\simeq \frac{e^{2}}{3\pi ^{2}}%
k_{\bot }^{2}\left[ \frac{1}{2}\ln \xi +\mathcal{O}(\xi ^{0})\right] +\nonumber\\
+\mathcal{O}(k_{\bot }^{4})\qquad\qquad\qquad\qquad
\end{eqnarray}
\begin{equation}
\Pi _{33}(k_{3}=0,k_{0}=0,k_{\bot }\sim 0)\simeq \frac{e^{2}}{a^{2}}\left[
\frac{1}{3}+\mathcal{O}(\xi ^{2})\right] +\mathcal{O}(k_{\bot }^{2})
\label{28a}
\end{equation}
where $\xi =am/2\pi \ll 1$ . Using the results (\ref{28}) and (\ref{28a}) in
the dispersion equations (\ref{13}), (\ref{14}), and taking into account
that the photon velocity for each propagation mode can be obtained from $%
\mathrm{v}(\mathbf{k})=\partial k_{0}/\partial \left|
\mathbf{k}\right| $, we find that within the considered
approximation the transverse and longitudinal modes propagate
perpendicularly to the compactified direction with velocities

\begin{equation}
\mathrm{v}_{\bot }^{T}\simeq 1-\frac{e^{2}}{12\pi ^{2}}\ln \xi ,  \label{18}
\end{equation}
\begin{equation}
\mathrm{v}_{\bot }^{L}\simeq 1-\left[ (M_{\bot }^{L})^{2}/2k_{\bot
}^{2}\right] ,  \label{18a}
\end{equation}
respectively, where $(M_{\bot }^{L})^{2}=\Pi _{33}=e^{2}/3a^{2}>0$ plays the
role of an effective topological mass for the longitudinal mode.

We call the reader's attention to the fact that the modifications found for the two velocities, $%
\mathrm{v}_{\bot }^{T}$ and $\mathrm{v}_{\bot }^{L}$, have
different origins. The modification of the longitudinal velocity
$\mathrm{v}_{\bot }^{L}$ is due to the appearance of the
topological mass $M_{\bot }^{L}$; while the transverse
superluminal velocity $\mathrm{v}_{\bot }^{T}$ (note that
$\mathrm{v}_{\bot }^{T}>c $ because $\xi <1$ in the used
approximation) appears as a consequence of a genuine variation of
the refraction index in the considered spacetime. Modifications of
the photon speed in non-trivial vacua have been previously
reported in the literature  \cite{Scharn1}.

It is easy to corroborate that for compactification lengths in
agreement with the used approximation, $a<1/m\sim 10^{3}\,fm$, the
transverse velocity (\ref{18}) is about 0.1\% larger than the
light
velocity in trivial spacetime. We underline that albeit $%
\mathrm{v}_{\bot }^{T}>c,$ there is no causality violation in this
problem. To understand this, let us recall that the velocity
(\ref{18}) is a low-frequency mode velocity. On the other hand,
the velocity of interest for signal propagation, and hence the
relevant one for causality, is the high-frequency velocity
$\mathrm{v}_{\bot }^{T}(q_{0}\rightarrow \infty )$. To determine
the difference between the two, one would need to investigate the
absorption coefficient, $\textit{Im}$$[n(q_{0})]$, with $n(q_{0})$
being the refraction index as a function of the frequency in the
space with $S^{1}\times R^{3}$ topology. However, aside from any
needed calculation, we agree with the analysis of Refs.
\cite{causality} regarding the lack of causality violations in
similar systems. We believe that in the case under study no
(micro-)causality should be violated, because the events taking
place in the $S^{1}\times R^{3}$ space are not constrained by the
null cone of a Minkowskian system, as Lorentz symmetry is
explicitly broken in the present situation.

The low-frequency limit ( $k_{0}=0$, $\left| \mathbf{k}\right| \rightarrow 0$%
) used to obtain the longitudinal-mode mass $M_{\bot }^{L}$ is
essential to study the static properties of the electromagnetic
field in this space. The mass obtained in this limit plays the
role of a magnetic mass of the longitudinal electromagnetic mode
\cite{Ferrer}. As showed in Ref. \cite {Romeo}, this topological
mass affects the magnetic response of the system.

Considering now photons propagating along the $\mathcal{OZ}$%
-direction ($k_{\perp }=0$), the dispersion relations (\ref{11}%
) and (\ref{12}) can be written respectively as

\begin{equation}
k_{0}^{2}-k_{3}^{2}-\Pi _{11}=0  \label{32}
\end{equation}

\begin{equation}
k_{0}^{2}-k_{3}^{2}-\frac{k_{0}^{2}-k_{3}^{2}}{k_{0}^{2}}\Pi _{33}=0
\label{32a}
\end{equation}

Assuming $ak_{0}\ll am\ll 1$ in (\ref{15}), the components of the
polarization operator appearing in (\ref{32}) and (\ref{32a}) become

\begin{equation}
\Pi _{11}(k_{\bot }=0,k_{3},k_{0}\sim 0)\simeq \frac{e^{2}}{a^{2}}\left[
\frac{1}{9}+\mathcal{O}(\xi ^{2})\right] +\mathcal{O}(k_{0}^{2})  \label{33}
\end{equation}
\begin{equation}
\Pi _{33}(k_{\bot }=0,k_{3},k_{0}\sim 0)\simeq \frac{e^{2}}{a^{2}}\frac{%
k_{0}^{2}}{k_{3}^{2}}\left[ \frac{-1}{9}+\mathcal{O}(\xi ^{2})\right] +%
\mathcal{O}(k_{0}^{4})  \label{33a}
\end{equation}

From (\ref{32}), (\ref{32a}), (\ref{33}) and (\ref{33a}) we can
straightforwardly find the low-frequency limit of the photon velocities for
transverse and longitudinal modes propagating along the $\mathcal{OZ}$%
-direction

\begin{equation}
\mathrm{v}_{\Vert }^{T}\simeq 1-\left[ (M_{\Vert
}^{T})^{2}/2k_{3}^{2}\right] ,  \label{34}
\end{equation}
\begin{equation}
\mathrm{v}_{\Vert }^{L}\simeq 1-\left[ (M_{\Vert
}^{L})^{2}/2k_{3}^{2}\right] ,  \label{34a}
\end{equation}
where both transverse and longitudinal mode masses coincide and are given by
$M_{\Vert }^{T}=M_{\Vert }^{L}=e/3a$. We stress that in this case both
velocities are smaller than the light velocity in trivial Minkowski space $%
c, $ and that the modification is due, as in (\ref{18a}), to the appearance
of a topological photon mass for each mode.

\section{Effect of Compactification on Fermion Condensation}

Dynamical chiral symmetry breaking in phenomenological models with
fermion interactions of Nambu-Jona-Lasinio (NJL) type have
attracted a great deal of attention \cite{DSB}, \cite{Kim} since
its introduction in the seminal paper of Nambu and Jona-Lasinio
\cite{Nambu}. Our interest here is to consider dynamical chiral
symmetry breaking in a non-trivial topological space taking into
account the results of Sec. II. With this aim, let us add a NJL
four-fermion term with coupling $G$ to massless QED, so that its
Lagrangian density becomes

\begin{equation}
\mathcal{L}=-\frac{1}{4}F_{\mu \nu }^{2}+\overline{\psi }(i\partial \llap / %
-eA\llap / )\psi +\frac{G}{2N}\left[ \left( \overline{\psi }\psi
\right) ^{2}+\left( \overline{\psi }i\gamma ^{5}\psi \right)
^{2}\right] . \label{NJL}
\end{equation}
The fermions in (\ref{NJL}) are assumed to carry out a flavor
index $\alpha =1,2,\dots ,N$. Introducing the composite
fields $\sigma =-\frac{G}{N}\left( \overline{\psi }\psi \right) ,$ $\pi =-%
\frac{G}{N}\left( \overline{\psi }i\gamma ^{5}\psi \right) $, the
gauged-NJL Lagrangian density (\ref{NJL}) can be rewritten as
\begin{equation}
\mathcal{L}_{NJL}=\frac{1}{2}\left[ \overline{\psi },i\gamma ^{\mu }D_{\mu
}\psi \right] -\overline{\psi }\left( \sigma +i\gamma ^{5}\pi \right) \psi -%
\frac{N}{2G}\left( \sigma ^{2}+\pi ^{2}\right) .  \label{NJL-1}
\end{equation}

The Lagrangian (\ref{NJL}) has a continuous chiral symmetry $\psi
\rightarrow e^{i\alpha \gamma ^{5}}\psi $, but it is clear from
Eq. (\ref{NJL-1}) that
if $\sigma $ gets a different from zero vacuum expectation value (vev) $\bar{%
\sigma}$,  this chiral symmetry is broken and the fermions acquire
mass. We are interested in the effective potential in the
large-$N$ limit, assuming $e\ll G$. The effective potential is a
function of the scalar and pseudo-scalar fields $\sigma $ and $\pi
$ respectively, and can be obtained by integrating out all the
fluctuation fields in the path integral. In a flat and
topologically trivial spacetime, the effective potential in
leading order at large-N, is given, after performing the Wick
rotation to Euclidean space, by

\begin{equation}
V(\sigma )=\frac{\sigma ^{2}}{2G}-2\int^{\Lambda
}\frac{d^{4}p_{E}}{(2\pi )^{4}}\ln (\sigma ^{2}+p_{E}^{2}),
\label{V}
\end{equation}
where $\Lambda $ is a large momentum cutoff. In expression
(\ref{V}) we dropped all the $\sigma$-independent terms, as they
will not contribute to the stationary solution of the potential.
We considered a configuration with $\pi =0$ and $\sigma $=constant, since the effective potential $V$ only depends on the chiral invariant $%
\rho ^{2}=\sigma ^{2}+\pi ^{2}$.

The vacuum solution is determined by the stationary point of the
effective potential (\ref{V}). At $G>G_{c}=4\pi ^{2}/\Lambda
^{2}$, the stationary equation $\partial V(\sigma )/\partial
\sigma =0$ has a non-trivial solution $\bar{\sigma}$ that
corresponds to a global minimum of (\ref {V}).

If the third spatial dimension is compactified in a circle of radius $%
a $, the potential (\ref{V}) for antiperiodic fermions becomes

\begin{eqnarray}\label{T-AP}
V_{AP}(\sigma )=\frac{\sigma ^{2}}{2G}-2\int^{\Lambda }\frac{d^{4}p_{E}}{%
(2\pi )^{4}}\ln (\sigma ^{2}+p_{E}^{2})-\nonumber\\-\frac{4}{a}\int \frac{d^{3}\widehat{p%
}}{(2\pi )^{3}}\ln (1+e^{-a\overline{\varepsilon }_{p}}),\qquad
\end{eqnarray}
with $\overline{\varepsilon }_{p}=\sqrt{\widehat{p}^{2}+\sigma
^{2}}$. The last term in the RHS of Eq. (\ref{T-AP}) is obtained
after summing in the discrete momentum $p_{3}$. The appearance of
this new term gives rise to the critical coupling
$G_{c}^{a}=6a^{2}G_{c}/(6a^{2}-G_{c})$, which now depends on the
compactification radius $a$. Notice that $G_{c}^{a}>G_{c}$. Hence,
for twisted fermions the compactification tends to restore the
symmetry, in agreement with results previously found in Refs.
\cite{Kim}, \cite{Hosaka} within a pure NJL theory (without gauge
fields). When periodic fermions are considered in the trivial
vacuum, the corresponding effective potential $V_{P}$, is similar
to (\ref{T-AP}), with the only change of a negative sign in front
of the exponential in the RHS of Eq. (\ref{T-AP}). This case was
also studied in Ref. \cite{Kim} in the context of a pure NJL
model. There, the analysis of the minimum of the potential
revealed that the effect of the compactified dimension is to
enhance the chiral condensation, i.e., to decrease the critical
value of the coupling.

The situation is different however for the gauged-NJL theory with
untwisted fermions. Here, in analogy with QED, the stable vacuum
is given by the constant vector potential (\ref{26a}).
Consequently, the effective potential must explicitly depend on a
nontrivial vacuum solution. Summing in the discrete momentum
$p_{3}$ and taking into account that the constant vacuum
(\ref{20}) enters in the calculation of the effective potential as
a shift $p_{3}\rightarrow p_{3}+\Delta$ in this discrete variable,
the effective potential results

\begin{eqnarray}\label{T-P}
V_{P}(\sigma ,\Delta )=\frac{\sigma ^{2}}{2G}-2\int^{\Lambda }\frac{%
d^{4}p_{E}}{(2\pi )^{4}}\ln (\sigma ^{2}+p_{E}^{2})-\nonumber\\-\frac{4}{a}\int \frac{%
d^{3}\widehat{p}}{(2\pi )^{3}}\ln \left[ 1-e^{-a\left(
\overline{\varepsilon }_{p}-ie\Delta \right) }\right].
\end{eqnarray}
Notice that when the minimum solution (\ref{26a}) is substituted
on (\ref{T-P}), one obtains $V_{P}(\sigma ,\Delta _{\min
})=V_{AP}(\sigma )$. As a consequence, the critical coupling for
the chiral condensation with untwisted fermions reduces to the
same $G_{c}^{a}$ already found for twisted fermions in the trivial
vacuum. Thus, we conclude that, independently of the fermion
boundary condition, the effect of the compactification in the
theory (\ref{NJL}) is to counteract the condensation, and
eventually, to reinstate the chiral symmetry at some critical
value of the compactification radius.

\section{Concluding Remarks}

In this paper we have shown that in a nonsimply connected
spacetime with topology $S^{1}\times R^{3}$, the stable vacuum
solution for QED with untwisted fermions is given by constant
field configurations that are gauge equivalent to
$\overline{A}_{3}=\frac{\pi }{ea}$, while for twisted fermions the
stable solutions correspond to constant gauge configurations
equivalent to the trivial vacuum. As a consequence, the one-loop
effective potentials for twisted and untwisted fermions coincide
when the corresponding stable vacuum solutions are considered. A
direct implication of the relation between the fermion boundary
conditions and the QED vacua is that the vacuum polarization
cannot distinguish between the two classes of fermions, once the
corresponding true vacuum is taking into account.

Another interesting outcome of this investigation is the
anisotropy in the photon propagation due to the nontrivial
topology of the spacetime. In the $S^{1}\times R^{3}$ domain, the
photons have several massive modes and a transverse superluminal
one. The masses of the photon modes increase as the inverse of the
compactification radius, while the superluminal velocity of the
massless mode increases logarithmically with that topological
distance. The existence of massive modes implies that at very
small radius of compactification, the photon propagation at low
energies is effectively confined to a Minkowskian
$(2+1)-$dimensional manifold, on which only superluminal photons
propagate. Therefore, photons moving in such a lower dimensional
space experience the lack of Lorentz symmetry of the general
manifold ($S^{1}\times R^{3}$) on which the lower-dimensional
space is embedded, allowing them to have a group velocity larger
than the usual Minkowskian velocity $c$.

We also considered how the non-trivial topology affects the
condensation of fermion-antifermion pairs. This was done in the
framework of QED with an additional four-fermion interaction
(gauged-NJL theory). In this model we found that the smaller the
compactification radius, the larger the critical four-fermion
coupling needed to generate a fermion-antifermion chiral symmetry
breaking condensate. Contrary to what occurs in a pure NJL model
\cite{Kim}, this result is obtained for both twisted and untwisted
fermions, once the corresponding stable vacuum is considered.
Thus, we conclude that in the gauged-NJL theory the
compactification tends to reinstate the chiral symmetry.

The results we are reporting here can be of interest for condensed
matter quasi-planar systems, as well as for theories with extra
dimensions.

\textbf{Acknowledgments}

This research was supported by the National Science Foundation
under Grant No. PHY-0070986.

\end{document}